\documentclass[10pt,letterpaper]{article}
\usepackage{spconf,amsmath,amssymb,graphicx,hyperref}
\usepackage[letterpaper,margin=1in,heightrounded]{geometry}
\usepackage{array}
\usepackage{booktabs}
\usepackage{makecell}
\usepackage{microtype}
\microtypesetup{protrusion=false}
\usepackage{multirow}
\usepackage{placeins}
\usepackage{tabularx}
\usepackage{xcolor}
\usepackage{url}
\hypersetup{hidelinks}

\newcolumntype{Y}{>{\centering\arraybackslash}X}
\renewcommand{\arraystretch}{0.96}
\makeatletter
\renewcommand\section{\@startsection{section}{1}{\z@}%
  {1.5ex plus .3ex minus .2ex}{0.8ex plus .2ex}{\normalfont\normalsize\bfseries}}
\renewcommand\subsection{\@startsection{subsection}{2}{\z@}%
  {1.0ex plus .2ex minus .1ex}{0.4ex plus .1ex}{\normalfont\normalsize\bfseries}}
\long\def\@makecaption#1#2{%
  \vskip 4pt
  \setbox\@tempboxa\hbox{#1. #2}%
  \ifdim \wd\@tempboxa >\hsize #1. #2\par
  \else \hbox to\hsize{\hfil\box\@tempboxa\hfil}\fi}

\def\thebibliography#1{\section{References}\fontsize{10pt}{12pt}\selectfont\list
 {\@biblabel{\arabic{enumi}}}{\settowidth\labelwidth{\@biblabel{#1}}%
 \leftmargin\labelwidth
 \advance\leftmargin\labelsep
 \setlength{\rightmargin}{2pt}
 \setlength{\itemsep}{0pt}%
 \setlength{\parsep}{0pt}%
 \setlength{\topsep}{0pt}%
 \setlength{\partopsep}{0pt}%
 \usecounter{enumi}}%
 \def\newblock{\hskip .11em plus .33em minus .07em}%
 \sloppy\clubpenalty10000\widowpenalty10000
 \sfcode`\.=1000\relax}

\makeatother

\title{PHONOLOGY-AWARE SPEAKER EXTRACTION FRONTEND FOR AUDIO LLMS}

\name{
\shortstack[c]{
Yanxiang Chen$^{1}$,
Haoyuan Deng$^{3}$,
Jinyu Wang$^{2}$,
Hongyi Chen$^{3}$\\
Dingshan Li$^{3}$,
Runlin Huang$^{3}$,
Weifeng Su$^{3}$
}
}

\address{
$^{1}$Sun Yat-Sen University, Zhuhai, China\\
$^{2}$Fudan University, Shanghai, China\\
$^{3}$Beijing Normal-Hong Kong Baptist University, Zhuhai, China
}
\newcommand{\tablefont}{\fontsize{10pt}{12pt}\selectfont}
\begin{document}
\maketitle
\raggedbottom

\begin{abstract}
Target speaker extraction (TSE) is a key frontend for enabling audio large language models (LLMs) to recognize and understand a designated speaker in overlapping speech.
However, multilingual mixtures exhibit heterogeneous extraction demands, causing fixed-depth systems to either under-process challenging inputs or waste computation on inputs that have already converged.
Existing TSE frontends generally apply uniform computation, while Audio-LLM adaptation methods neither explicitly model language-conditioned extraction demands nor preserve plug-and-play compatibility with frozen downstream models.
We therefore propose LATTE, a lightweight 5.2M-parameter frontend that combines an acoustic Difficulty Estimator, a task-adaptive phonology-informed prior learned from multi-depth TSE gains, and an Adaptive Hourglass Extractor to route each mixture without requiring language identity at inference.
On PolyExtract, LATTE achieves a macro SI-SNRi of 15.07\,dB at 0.30\,GFLOPs, retaining performance within 0.09\,dB of a 0.53-GFLOP fixed-depth model while reducing computation by 43.4\%, and lowers the average WER from 16.8\% to 14.6\% relative to TF-GridNet.
By jointly modeling sample-level extraction demands and phonological structure, LATTE provides a practical framework for quality-aware, compute-efficient multilingual TSE and offers a scalable path toward robust Audio-LLM deployment.
\end{abstract}


\begin{figure*}[!t]
  \centering
  \includegraphics[width=0.9\textwidth]{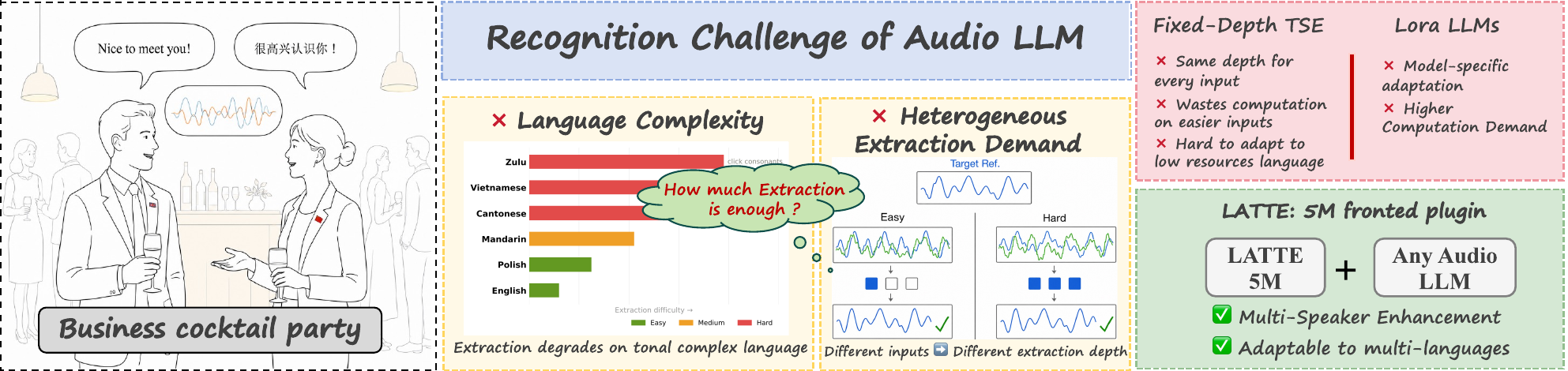}
  \caption{Motivation of the proposed LATTE framework.}
  \label{fig:motivation}
\end{figure*}

\begin{figure*}[t]
  \centering
  \includegraphics[width=0.9\textwidth]{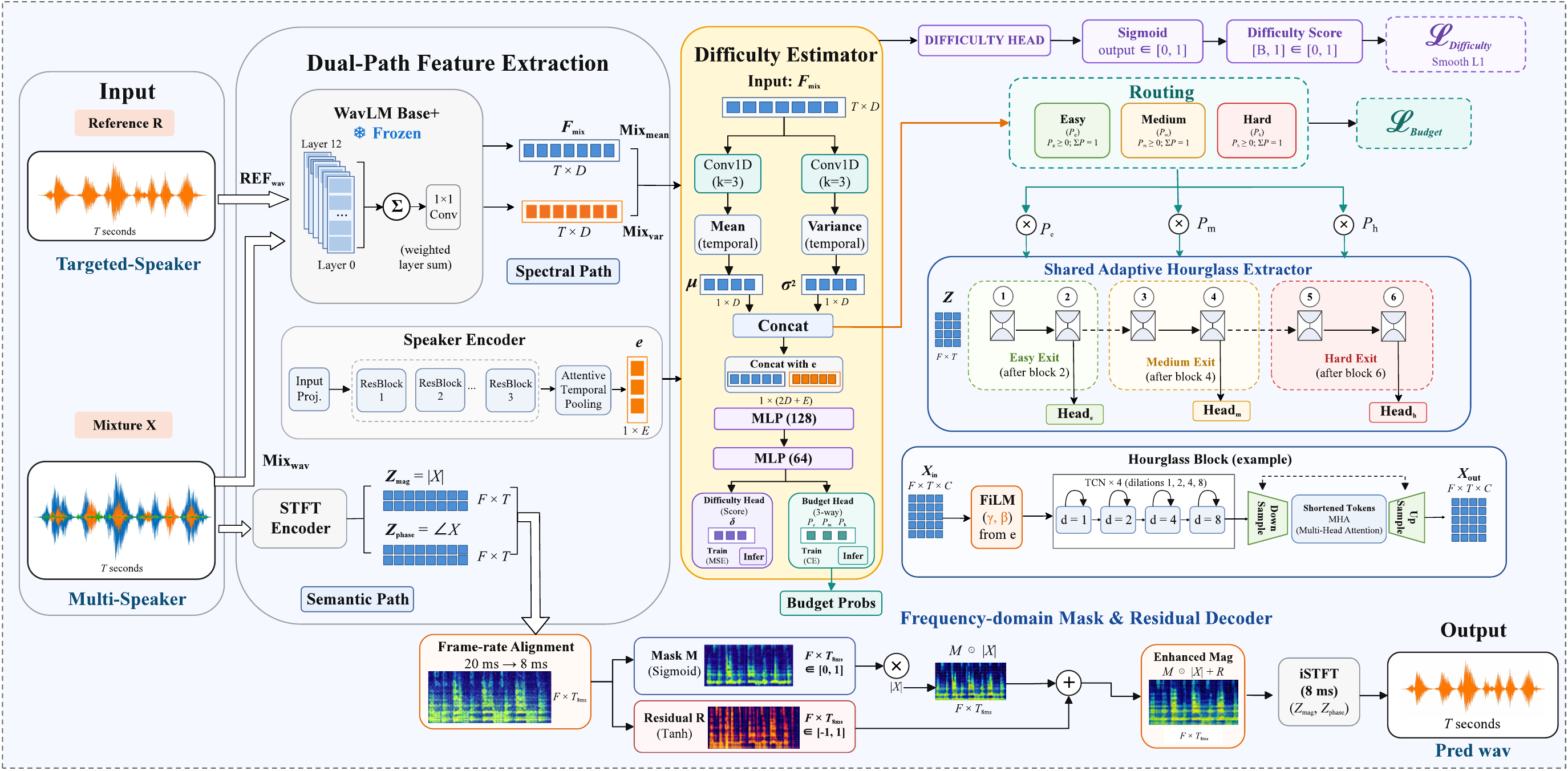}
  \caption{LATTE combines semantic and spectral evidence, estimates a
  low/medium/high compute budget, and routes the mixture through shared
  multi-exit hourglass blocks before target-speech reconstruction.}
  \label{fig:framework}
\end{figure*}

\section{Introduction}

\label{sec:intro}

Target speaker extraction (TSE) recovers a designated speaker from an
overlapping mixture using a short enrollment utterance. Existing systems
typically apply the same network depth to every input
~\cite{delcroix2018speakerbeam,ge2020spexplus,wang2023tfgridnet}. Our study begins with an overlooked empirical observation: under matched mixture conditions, speech from different languages can require substantially different amounts of computation to reach a common extraction-quality target. Controlled multi-depth experiments reveal, however, that multilingual
conditions can require markedly different amounts of computation to approach
their best extraction quality. A shallow budget under-processes demanding
inputs, whereas a deep budget repeatedly refines inputs that have already
converged. This motivates our central question: \emph{can a TSE system estimate
the demand of each input and execute only the computation needed for reliable
extraction?}

\textbf{Why can extraction demand vary across languages?} Cross-linguistic variation in tone, phonation, rhythm, and segmental structure
changes the acoustic organization of linguistic content and speaker cues
~\cite{gordon2001phonation,ramus1999correlates}. Because these factors interact
with mixture acoustics and target identity, we treat language structure as a
training prior rather than a deterministic or intrinsic measure of language
difficulty. Specifically, we derive a task-adaptive
\textbf{Phonological Difficulty Index (PDI)} from the minimum processing depth
needed to approach deepest-exit quality. URIEL+ and PHOIBLE
~\cite{khan2025uriel,moran2014phoible} provide a phonological similarity graph
that smooths this task-observed evidence across related training languages.

We propose \textbf{LATTE}, a phonology-aware adaptive TSE model evaluated on
\textbf{PolyExtract}, a controlled benchmark covering eight languages and 64
directed target--interferer conditions. LATTE combines a sample-level
Difficulty Estimator with a shared multi-exit Hourglass Extractor. The
estimator predicts an easy, medium, or hard computation budget from mixture
features and the target-speaker embedding, routing each input through two,
four, or six shared blocks. PDI supplies weak supervision only during training;
inference requires neither language identity nor access to an external
linguistic database.

LATTE also serves as a lightweight waveform frontend for frozen audio-language
backends, avoiding backend-specific adaptation such as LoRA
~\cite{hu2022lora}. On PolyExtract, it achieves 15.07\,dB macro SI-SNRi at
0.30\,GFLOPs, remaining within 0.09\,dB of the deepest fixed path while     
reducing computation by 43.4\%. With frozen Qwen3-Omni, it further reduces
average ASR WER from 16.8\% for TF-GridNet to 14.6\%. Our Contributions are:

\begin{enumerate}
    \item We identify heterogeneous compute demand as a central
quality--efficiency challenge in multilingual TSE;
    \item We introduce PolyExtract
and a task-adaptive, phonologically structured PDI for supervising demand
estimation;
    \item We develop LATTE, which performs language-label-free adaptive
routing and improves signal-level, computational, and downstream recognition
performance.
\end{enumerate}

\section{Method}
\label{sec:method}

\subsection{Architecture and adaptive extraction}
Given a mixture $\mathbf{x}\in\mathbb{R}^{T}$ and an enrollment utterance
$\mathbf{r}\in\mathbb{R}^{T'}$, LATTE estimates the target waveform
$\hat{\mathbf{s}}$. The frozen WavLM Base+ encoder and a shared learned
projection produce mixture and enrollment representations
\begin{equation}
\mathbf{F}_{\mathrm{mix}}
 = f_{\mathrm{proj}}\!\left(f_{\mathrm{wav}}(\mathbf{x})\right),
\qquad
\mathbf{F}_{\mathrm{ref}}
 = f_{\mathrm{proj}}\!\left(f_{\mathrm{wav}}(\mathbf{r})\right),
\end{equation}
where $\mathbf{F}_{\mathrm{mix}}\in\mathbb{R}^{D\times L_s}$ is the
initial extractor representation,
\begin{equation}
\mathbf{Z}_0=\mathbf{F}_{\mathrm{mix}},
\qquad
\mathbf{e}=\operatorname{AttPool}(\mathbf{F}_{\mathrm{ref}}).
\label{eq:extractor-input}
\end{equation}
Thus, the Hourglass extractor operates on the projected semantic mixture
features, while the target-speaker embedding $\mathbf{e}$ conditions every
block through FiLM. In parallel, a 512-point STFT with a 128-sample hop
produces the mixture magnitude $|\mathbf{X}|$ and phase
$\angle\mathbf{X}$. The STFT representation does not enter the Hourglass
extractor: after routing, the selected exit $\mathbf{o}$ is linearly
interpolated from the WavLM frame grid $L_s$ to the STFT frame grid $L_f$,
and the decoder predicts a mask $\mathbf{M}$ and bounded residual
$\mathbf{R}$ from the aligned exit and $|\mathbf{X}|$:
\begin{equation}
\widetilde{\mathbf{o}}
 = \operatorname{Align}_{L_s\rightarrow L_f}(\mathbf{o}),
\qquad
[\mathbf{M},\mathbf{R}]
 = f_{\mathrm{dec}}([\widetilde{\mathbf{o}};|\mathbf{X}|]).
\end{equation}
The enhanced magnitude is
$|\hat{\mathbf{S}}|=\max\{0,\mathbf{M}\odot|\mathbf{X}|+\mathbf{R}\}$,
and the waveform is reconstructed as
$\hat{\mathbf{s}}=\operatorname{iSTFT}
(|\hat{\mathbf{S}}|,\angle\mathbf{X})$.

The difficulty estimator (DE) summarizes the mean and variance of mixture
features $\mathbf{F}$, fuses them with $\mathbf{e}$, and predicts demand
$\delta\in[0,1]$ and one-hot budget $\mathbf{b}\in\{0,1\}^{3}$. Training uses
straight-through Gumbel-Softmax~\cite{jang2017categorical}; inference uses
$\arg\max$. Six hourglass blocks form three nested tiers, with exits after
blocks 2, 4, and 6. Each block injects $\mathbf{e}$ through FiLM
conditioning~\cite{perez2018film}, captures local structure with dilated
depthwise TCNs~\cite{luo2019convtasnet}, and models global context by attention
on a downsampled sequence~\cite{subakan2021sepformer}. If $\phi_k$ is exit
$k$ and $\mathbf{h}_{2k}$ its hidden state, the routed representation is
\begin{equation}
  \mathbf{o}=\sum_{k=1}^{3}b_k\,\phi_k(\mathbf{h}_{2k}).
  \label{eq:exit}
\end{equation}
The tiers share prefixes rather than replicate experts; early exiting therefore
reduces executed depth while keeping the extractor at 2.7M parameters.

\subsection{Task-adaptive phonological difficulty index (PDI)}
We define PDI as a target-language scalar $d_\ell=\sigma(a_\ell)\in[0,1]$:
the normalized exit depth required under the current model, data, and
tolerance; low/high values mean shallow/deep demand. If $Q_{i,k}$ is sample $i$'s
quality at exit $k$, its task evidence is the detached minimum-depth target
\begin{equation}
 \begin{aligned}
 k_i^*&=\min\{k:Q_{i,3}-Q_{i,k}\leq\epsilon\},\\
 \widetilde y_i&=\operatorname{sg}\!\left(\frac{k_i^*-1}{2}\right).
 \end{aligned}
 \label{eq:demand}
\end{equation}
The stop-gradient prevents a trivial solution that degrades shallow exits to
lower their targets. The language-balanced mean $\bar y_\ell$ anchors $d_\ell$,
which is smoothed on a frozen URIEL+/PHOIBLE similarity graph $S$. With Smooth
L1 loss $\rho$,
\begin{align}
 \mathcal{L}_{\rm PDI}&=\tfrac1{|\mathcal G|}\sum_\ell
 \rho(d_\ell-\operatorname{sg}(\bar y_\ell)) \nonumber\\
 &\quad+\tfrac{\lambda_g}{Z}\sum_{\ell<m}S_{\ell m}(d_\ell-d_m)^2, \nonumber\\
 \mathcal{L}_{\rm DE}&=\tfrac1N\sum_i\!\left[
 \rho(\delta_i-\widetilde y_i)+\lambda_l
 \rho(\delta_i-\operatorname{sg}(d_{\ell_i}))\right].
 \label{eq:pdi}
\end{align}
The graph only smooths task evidence; PDI is not accessed at test time.

\subsection{Training objective and inference}
Let $p_{i,k}$ be the soft routing probability, $C_k$ the cost of path $k$, and
$\mathcal{L}^{\rm ext}_{i,k}$ its extraction loss. LATTE minimizes
\begin{align}
 \mathcal{L}_{\rm route}
 &=\tfrac1N\sum_{i,k}p_{i,k}
 \bigl(\mathcal{L}^{\rm ext}_{i,k}+\lambda_C C_k\bigr), \nonumber\\
 \mathcal{L}
 &=\mathcal{L}_{\rm route}+\alpha\mathcal{L}_{\rm spk}
 +\beta\mathcal{L}_{\rm DE}+\gamma\mathcal{L}_{\rm PDI}.
 \label{eq:total-loss}
\end{align}
Here $\mathcal{L}^{\rm ext}$ is per-exit negative SI-SNR and
$\mathcal{L}_{\rm spk}$ regularizes the enrollment embedding; the cost term
discourages collapse to the deepest path. Training warms and directly
supervises all exits under uniform path sampling, freezes the phonological
graph, and derives detached targets from multi-exit quality curves. Joint
training anneals the Gumbel temperature, with loss weights and thresholds
selected on validation data. At inference, the DE runs once and only the
selected shared prefix and exit are executed, yielding conditional computation
without language identity.

\section{Experiments}
\label{sec:experiments}

\subsection{Experimental setup and PolyExtract}
LATTE has 5.2M parameters and is trained with Adam ($10^{-3}$ learning rate,
$10^{-5}$ weight decay), mixed precision, and gradient clipping. Training uses
online 4-s mixtures at $[-5,5]$ dB; validation and test mixtures are fixed.
All signals are resampled to 16 kHz, and the STFT uses a 512-point FFT with a
128-sample hop. PolyExtract contains eight languages from five families,
drawn from LibriSpeech, AISHELL-1, NCHLT, VIVOS, and Common Voice
~\cite{panayotov2015librispeech,bu2017aishell,barnard2014nchlt,
luong2016vivos,ardila2020common}. We evaluate all 64 directed
target--interferer pairs with equal sample counts and matched mixture controls.

\begin{table}[t]
\centering
\tablefont
\setlength{\tabcolsep}{3pt}
\begin{tabularx}{\columnwidth}{@{}l >{\raggedright\arraybackslash}X r@{}}
\toprule
\textbf{Language} & \textbf{Key Phonological Features} & \textbf{PDI} \\
\midrule
English         & non-tonal, moderate clusters             & 0.15 \\
Turkish         & agglutinative, 8-way vowel harmony      & 0.30 \\
Mandarin        & 4 tones, simple onsets                  & 0.35 \\
Polish          & complex clusters (\textit{szcz-}, \textit{prz-}) & 0.42 \\
Kinyarwanda     & 2 Bantu tones, prenasalized stops       & 0.48 \\
Cantonese       & 6--9 tones, checked codas               & 0.55 \\
Vietnamese      & 6 tones + phonation contrasts           & 0.65 \\
Zulu            & 3 click series, tones, prenasalized     & 0.72 \\
\bottomrule
\end{tabularx}

\caption{PolyExtract language overview.}
\label{tab:polyextract}
\end{table}

\subsection{Signal-Level Extraction across Language Pairs}
Table~\ref{tab:signal} reports pair-macro results on the same PolyExtract
split. Fixed-budget systems remain strong on low-demand conditions, whereas
LATTE gives the best cross-language, high-PDI, and worst-pair results at the
lowest average computation.

\begin{table*}[!t]
\centering
\tablefont
\begingroup
\setlength{\tabcolsep}{3pt}

\begin{tabular*}{\textwidth}{@{\extracolsep{\fill}}lcccccc@{}}
\toprule
& SpEx+
& \shortstack[c]{TD-\\SB}
& \shortstack[c]{AD-Flow\\TSE}
& TCE
& TF-GridNet
& \textbf{LATTE} \\
\midrule
\multicolumn{7}{l}{\textbf{Model Complexity}} \\
Trainable Params$\downarrow$ & 11.1M & 14.4M & 85.0M & \underline{8.7M} & 14.5M & \textbf{5.2M} \\
Avg. GFLOPs$\downarrow$ & 0.62 & 0.74 & 5.80 & \underline{0.55} & 1.42 & \textbf{0.30} \\
\midrule
\multicolumn{7}{l}{\textbf{All Directed Language Pairs}} \\
Macro SI-SNRi$\uparrow$ & 12.8 & 13.2 & \underline{14.6} & 13.5 & 13.8 & \textbf{15.07} \\
Macro PESQ$\uparrow$   & 2.92 & 2.98 & \underline{3.14} & 3.01 & 3.06 & \textbf{3.22} \\
Macro STOI$\uparrow$   & 0.87 & 0.88 & \underline{0.91} & 0.89 & 0.90 & \textbf{0.92} \\
\addlinespace[1pt]
\multicolumn{7}{l}{\textbf{Same- vs. Cross-Language Pairs}} \\
Same-language SI-SNRi$\uparrow$ & 14.9 & 15.3 & \textbf{16.1} & 15.5 & 15.6 & \underline{15.7} \\
Cross-language SI-SNRi$\uparrow$ & 12.5 & 12.9 & \underline{14.4} & 13.2 & 13.5 & \textbf{14.9} \\
\addlinespace[1pt]
\multicolumn{7}{l}{\textbf{Cross-Language PDI Bands}} \\
Low$\rightarrow$Low SI-SNRi$\uparrow$ & 16.0 & 16.3 & \textbf{16.8} & 16.4 & \underline{16.5} & 16.0 \\
Low$\rightarrow$High SI-SNRi$\uparrow$ & 13.2 & 13.6 & \underline{14.8} & 13.9 & 14.0 & \textbf{15.1} \\
High$\rightarrow$Low SI-SNRi$\uparrow$ & 11.7 & 12.1 & \underline{13.9} & 12.5 & 12.8 & \textbf{14.6} \\
High$\rightarrow$High SI-SNRi$\uparrow$ & 11.0 & 11.5 & \underline{13.3} & 11.9 & 12.3 & \textbf{14.3} \\
\addlinespace[1pt]
\multicolumn{7}{l}{\textbf{Robustness}} \\
Worst-pair SI-SNRi$\uparrow$ & 9.8 & 10.2 & \underline{12.4} & 10.7 & 11.1 & \textbf{14.0} \\
\bottomrule
\end{tabular*}
\endgroup

\caption{Signal-level extraction quality and computational cost across directed target--interferer language pairs. Results are macro-averaged over language pairs.}
\label{tab:signal}
\end{table*}

\subsection{Multi-Speaker ASR with Diverse Frontends}
We compare LoRA adaptation with plug-in extraction before frozen Qwen3-Omni.
LATTE obtains 14.6\% average WER without updating the backend, and ranks first
on five of eight languages while requiring only 0.2T--0.4T FLOPs.
LATTE leads on Cantonese, Vietnamese, Polish, Zulu, and Kinyarwanda, where
adaptive extraction better handles interference than a uniformly processed
frontend.

\begin{table*}[t]
\centering
\tablefont
\setlength{\tabcolsep}{3pt}

\begin{tabular*}{\textwidth}{@{\extracolsep{\fill}}llccccc@{}}
\toprule
\multirow{2}{*}{Language}
& \multirow{2}{*}{Metric}
& \multicolumn{3}{c}{LoRA}
& \multicolumn{2}{c}{Plugin} \\
\cmidrule(lr){3-5}
\cmidrule(lr){6-7}
&
& \shortstack[c]{Qwen3\\Omni}
& \shortstack[c]{Whisper\\v3}
& \shortstack[c]{DeSTA\\2.5-A}
& \shortstack[c]{Voice\\Filter}
& LATTE \\
\midrule
\multirow{2}{*}{PolyExtract}
& WER$\downarrow$   & \underline{15.4} & 18.8 & 15.6 & 20.6 & \textbf{14.6} \\
& FLOPs$\downarrow$ & 18.2T & 4.8T & 12.6T & \underline{0.6T} & \textbf{0.3T} \\
\midrule
\multirow{2}{*}{Chinese}
& CER$\downarrow$   & \textbf{11.8} & 13.5 & 12.9 & 17.8 & \underline{12.6} \\
& FLOPs$\downarrow$ & 18.2T & 4.8T & 12.6T & \underline{0.6T} & \textbf{0.3T} \\
\multirow{2}{*}{English}
& WER$\downarrow$   & \textbf{7.9} & 8.9 & \underline{8.1} & 12.3 & 8.2 \\
& FLOPs$\downarrow$ & 18.2T & 4.8T & 12.6T & \underline{0.6T} & \textbf{0.2T} \\
\multirow{2}{*}{Cantonese}
& CER$\downarrow$   & \underline{14.9} & 16.4 & 21.5 & 32.6 & \textbf{13.8} \\
& FLOPs$\downarrow$ & 18.2T & 4.8T & 12.6T & \underline{0.6T} & \textbf{0.4T} \\
\multirow{2}{*}{Turkish}
& WER$\downarrow$   & 11.8 & \textbf{9.5} & 14.2 & 20.8 & \underline{9.8} \\
& FLOPs$\downarrow$ & 18.2T & 4.8T & 12.6T & \underline{0.6T} & \textbf{0.3T} \\
\multirow{2}{*}{Vietnamese}
& WER$\downarrow$   & \underline{16.8} & 17.1 & 26.3 & 34.2 & \textbf{14.2} \\
& FLOPs$\downarrow$ & 18.2T &  4.8T & 12.6T & \underline{0.6T} & \textbf{0.4T} \\
\multirow{2}{*}{Polish}
& WER$\downarrow$   & 13.8 & \underline{11.6} & 16.1 & 23.7 & \textbf{11.0} \\
& FLOPs$\downarrow$ & 18.2T & 4.8T & 12.6T & \underline{0.6T} & \textbf{0.3T} \\
\multirow{2}{*}{Zulu}
& WER$\downarrow$   & \underline{20.1} & 21.3 & 29.8 & 38.5 & \textbf{15.5} \\
& FLOPs$\downarrow$ & 18.2T & 4.8T & 12.6T & \underline{0.6T} & \textbf{0.4T} \\
\multirow{2}{*}{Kinyarwanda}
& WER$\downarrow$   & 16.4 & \underline{15.2} & 27.1 & 30.4 & \textbf{12.6} \\
& FLOPs$\downarrow$ & 18.2T & 4.8T & 12.6T & \underline{0.6T} & \textbf{0.3T} \\
\bottomrule
\end{tabular*}

\caption{Multilingual ASR performance across speaker extraction approaches.}
\label{tab:multilingual_performance}
\end{table*}

\begin{table*}[t]
\centering
\tablefont
\setlength{\tabcolsep}{2.4pt}
\begin{tabular*}{\textwidth}{@{\extracolsep{\fill}}ll r rrr rrr rrr@{}}
\toprule
& &
& \multicolumn{3}{c}{\textbf{PolyExtract (Avg.)}}
& \multicolumn{3}{c}{\textbf{Mandarin}}
& \multicolumn{3}{c}{\textbf{English}} \\
\cmidrule(lr){4-6}
\cmidrule(lr){7-9}
\cmidrule(lr){10-12}
\textbf{Category}
& \textbf{Method}
& \shortstack{\textbf{Trainable}\\\textbf{Params}$\downarrow$}
& \shortstack{Mix\\WER}
& \shortstack{Ext.\\WER$\downarrow$}
& RelRed
& \shortstack{Mix\\CER}
& \shortstack{Ext.\\CER$\downarrow$}
& RelRed
& \shortstack{Mix\\WER}
& \shortstack{Ext.\\WER$\downarrow$}
& RelRed \\
\midrule
\multirow{3}{*}{LoRA}
& Qwen3-Omni~\cite{xu2025qwen3omni}
& 7.0B
& 38.5 & \underline{15.4} & \textbf{60.0\%}
& 32.4 & \textbf{11.8} & \textbf{63.6\%}
& 24.6 & \textbf{7.9} & \textbf{67.9\%} \\
& Whisper-v3~\cite{openai2023whisperlargev3}
& 1.5B
& 38.5 & 18.8 & 51.2\%
& 32.4 & 13.5 & 58.3\%
& 24.6 & 8.9 & 63.8\% \\
& DeSTA2.5-Audio~\cite{lu2025desta25audio}
& 7.2B
& 38.5 & 15.6 & 59.5\%
& 32.4 & 12.9 & 60.2\%
& 24.6 & \underline{8.1} & 67.1\% \\
\midrule
\multirow{5}{*}{Plug-in}
& VoiceFilter~\cite{maum2020voicefilter}
& \underline{8.2M}
& 38.5 & 20.6 & 46.5\%
& 32.4 & 17.8 & 45.1\%
& 24.6 & 12.3 & 50.0\% \\
& TD-SpeakerBeam~\cite{delcroix2020speakerbeam}
& 14.4M
& 38.5 & 19.8 & 48.6\%
& 32.4 & 16.9 & 47.8\%
& 24.6 & 11.4 & 53.7\% \\
& SpEx+~\cite{ge2020spexplus}
& 11.1M
& 38.5 & 18.4 & 52.2\%
& 32.4 & 15.6 & 51.9\%
& 24.6 & 10.3 & 58.1\% \\
& TF-GridNet~\cite{wang2023tfgridnet}
& 14.5M
& 38.5 & 16.8 & 56.4\%
& 32.4 & 14.1 & 56.5\%
& 24.6 & 9.4 & 61.8\% \\
& \textbf{LATTE (ours)}
& \textbf{5.2M}
& 38.5 & \textbf{14.6} & \underline{62.1\%}
& 32.4 & \underline{12.6} & \underline{61.1\%}
& 24.6 & 8.2 & \underline{66.7\%} \\
\midrule
\multicolumn{2}{l}{\textit{Oracle (clean, single-spk)}}
& ---
& \multicolumn{3}{c}{8.3}
& \multicolumn{3}{c}{6.5}
& \multicolumn{3}{c}{4.1} \\
\bottomrule
\end{tabular*}%

\caption{Multi-speaker ASR results on PolyExtract.}
\label{tab:asr}
\end{table*}

\begin{table*}[t]
\centering
\tablefont
\setlength{\tabcolsep}{1.5pt}
\renewcommand{\arraystretch}{0.96}

\begin{tabularx}{\linewidth}{
  @{}
  >{\raggedright\arraybackslash}p{0.19\linewidth}
  *{12}{Y}
  @{}
}
\toprule
& \multicolumn{4}{c}{\shortstack[c]{Task A:Slot Filling}}
& \multicolumn{4}{c}{\shortstack[c]{Task B:Yes/No}}
& \multicolumn{4}{c}{\shortstack[c]{Task C:Attribution}} \\
\cmidrule(lr){2-5}
\cmidrule(lr){6-9}
\cmidrule(lr){10-13}
System
& en & zh & yue & Poly.
& en & zh & yue & Poly.
& en & zh & yue & Poly. \\
\midrule
Mix$\to$Qwen
& 24 & 21 & 15 & 20
& 61 & 58 & 51 & 57
& 43 & 40 & 33 & 39 \\
LoRA$\to$Qwen
& 70 & 57 & 47 & 58
& 88 & 77 & 67 & 77
& 85 & 70 & 59 & 71 \\
\textbf{LATTE$\to$Qwen}
& \textbf{58} & \textbf{55} & \textbf{49} & \textbf{54}
& \textbf{78} & \textbf{75} & \textbf{69} & \textbf{74}
& \textbf{72} & \textbf{68} & \textbf{61} & \textbf{67} \\
\textit{Clean$\to$Qwen}
& \textit{74} & \textit{69} & \textit{61} & \textit{68}
& \textit{91} & \textit{89} & \textit{83} & \textit{88}
& \textit{89} & \textit{85} & \textit{79} & \textit{84} \\
\bottomrule
\end{tabularx}
\par\vspace{6pt}
\begin{tabularx}{\linewidth}{
  @{}
  >{\raggedright\arraybackslash}p{0.19\linewidth}
  *{8}{Y}
  @{}
}
\toprule
& \multicolumn{4}{c}{\shortstack[c]{Task D:Intent Inference}}
& \multicolumn{4}{c}{\shortstack[c]{Task E:Emotion Recognition}} \\
\cmidrule(lr){2-5}
\cmidrule(lr){6-9}
System
& en & zh & yue & Poly.
& en & zh & yue & Poly. \\
\midrule
Mix$\to$Qwen
& 31 & 27 & 19 & 26
& 36 & 32 & 25 & 31 \\
LoRA$\to$Qwen
& 79 & 64 & 53 & 65
& 75 & 64 & 55 & 65 \\
\textbf{LATTE$\to$Qwen}
& \textbf{66} & \textbf{62} & \textbf{55} & \textbf{61}
& \textbf{67} & \textbf{62} & \textbf{53} & \textbf{61} \\
\textit{Clean$\to$Qwen}
& \textit{83} & \textit{78} & \textit{69} & \textit{77}
& \textit{81} & \textit{76} & \textit{68} & \textit{75} \\
\bottomrule
\end{tabularx}

\caption{LLM semantic understanding accuracy (\%).}
\label{tab:semantic}
\end{table*}

\begin{table}[t]
\centering
\tablefont
\setlength{\tabcolsep}{2pt}
\begin{tabularx}{\columnwidth}{@{}>{\raggedright\arraybackslash}X c c c@{}}
\toprule
\textbf{Variant} & \textbf{Avg} & \textbf{Hard} & \textbf{GFLOPs} \\
\midrule
\textbf{LATTE-Full} & \textbf{15.1} & \textbf{14.6} & \textbf{0.30g} \\
\midrule
\multicolumn{4}{@{}l}{\textit{Adaptive compute}} \\
W/o DE (Fixed-6)                 & 13.5 & 11.8 & 0.53g \\
W/o refinement                   & 14.3 & 12.5 & 0.28g \\
W/o $\mathcal{L}_{\textsc{bgt}}$ & 14.9 & 13.0 & 0.39g \\
\midrule
\multicolumn{4}{@{}l}{\textit{DE design}} \\
DE: SNR only                     & 13.9 & 12.2 & 0.32g \\
DE: LangID only                  & 14.2 & 12.8 & 0.32g \\
\midrule
\multicolumn{4}{@{}l}{\textit{Encoder/decoder}} \\
W/o MS encoder                   & 13.8 & 12.6 & 0.30g \\
W/o bound. residual $R$          & 14.2 & 12.3 & 0.30g \\
Semantic only (w/o spectral)     & 13.3 & 11.5 & 0.26g \\
Spectral only (w/o WavLM)        & 12.7 & 10.9 & 0.29g \\
\bottomrule
\end{tabularx}

\caption{Ablation study on PolyExtract. \textbf{Hard}: average over yue/vi/zu.
GFLOPs are per-utterance averages over the test set.}
\label{tab:ablation}
\end{table}

\subsection{Downstream Semantic Understanding}
We pass extracted waveforms to frozen Qwen3-Omni and evaluate slot filling,
yes/no comprehension, speaker attribution, intent inference, and emotion
recognition. LATTE consistently improves over direct mixture input while
leaving the Audio LLM unchanged.
With the backend frozen across conditions, gains over mixture input reflect
cleaner frontend evidence rather than task-specific Audio-LLM adaptation.

\subsection{Difficulty Estimation and Adaptive Computation}
\begin{figure*}[!t]
    \centering
    
    \includegraphics[width=0.48\textwidth,viewport=0 166 226.83 315.95,clip]{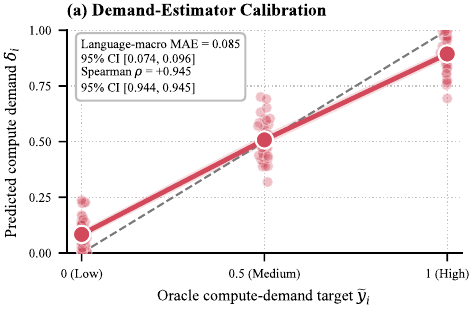}%
    \hfill
    \raisebox{3bp}{\includegraphics[width=0.48\textwidth,viewport=0 0 226.83 149.95,clip]{pdi_routing_evaluation.pdf}}
    \caption{Evaluation of difficulty estimation and adaptive computation.
    (a) Calibration of the predicted compute demand against the task-observed
    oracle target. Large markers denote demand-level means, and the dashed line
    denotes perfect calibration.
    (b) Quality--computation comparison among fixed-depth and adaptive routing
    strategies. The upper-left region represents higher extraction quality with
    lower average computation.}
    \label{fig:routing_evaluation}
\end{figure*}

Figure~\ref{fig:routing_evaluation} evaluates calibration and the
quality--computation frontier. The DE obtains language-macro MAE 0.085 and
Spearman $\rho=0.945$. LATTE reaches 15.07 dB at 0.30 GFLOPs, within 0.09 dB
of the deepest path while reducing computation by 43.4\%; learned-only,
task-PDI-only, random, and oracle routing obtain 14.69, 14.88, 13.91, and
15.23 dB, respectively.
The 0.16-dB oracle gap shows that most of the quality--compute trade-off is
recovered without labels, while the single-source variants confirm
complementary routing evidence.

\subsection{Ablation Study}
Table~\ref{tab:ablation} retains the complete component study. Difficulty
estimation and iterative refinement provide the largest gains on demanding
languages, while both representation paths and decoder components contribute.
Degradation from the semantic-only and spectral-only variants confirms that
WavLM context and fine-grained spectral evidence play complementary roles in
extraction and reconstruction.

\section{Conclusion}
LATTE is a lightweight adaptive TSE frontend for frozen Audio LLMs. PDI captures model-specific language-level compute demand from multi-depth evidence, while routing remains language-label free at inference. It is task-adaptive and improves routing over task-PDI-only ($15.07$ vs.\ $14.88$\,dB at $0.30$ GFLOPs).
Overall, task-observed multi-exit evidence and sample acoustics enable
near-oracle, language-label-free routing through one shared extractor, while
its waveform interface remains compatible with frozen downstream backends.
The shared exits turn the estimated demand into an explicit computational
budget: easier mixtures stop after a short processing prefix, whereas harder
mixtures receive additional refinement without maintaining separate
language-specific experts. The component study further shows that WavLM
context and fine-grained spectral detail provide complementary evidence for
extraction and reconstruction. Together with waveform-level output, this
separation allows one adaptive frontend to support different frozen downstream
systems while leaving their parameters unchanged.
PolyExtract covers eight languages and controlled two-speaker mixtures, so
generalization to broader language, accent, and real-world overlap conditions
remains unverified. Future work should evaluate unseen languages, accents, and
real-world overlap, together with deployment latency and energy.

\section{Compliance with Ethical Standards}
This study used only publicly available speech datasets in accordance with their respective licenses and terms of use. No new human participants were recruited, and no new personal data were collected. 

\section{Acknowledgments}
No external funding was received for this study. The authors have no relevant financial or non-financial interests to disclose.
\bibliographystyle{IEEEbib}
\bibliography{references}

@article{xu2025qwen3omni,
  title = {Qwen3-Omni Technical Report},
  author = {Xu, Jin and Guo, Zhifang and Hu, Hangrui and Chu, Yunfei and Wang, Xiong and He, Jinzheng and Wang, Yuxuan and Shi, Xian and He, Ting and Zhu, Xinfa and Lv, Yuanjun and Wang, Yongqi and Guo, Dake and Wang, He and Ma, Linhan and Zhang, Pei and Zhang, Xinyu and Hao, Hongkun and Guo, Zishan and Yang, Baosong and Zhang, Bin and Ma, Ziyang and Wei, Xipin and Bai, Shuai and Chen, Keqin and Liu, Xuejing and Wang, Peng and Yang, Mingkun and Liu, Dayiheng and Ren, Xingzhang and Zheng, Bo and Men, Rui and Zhou, Fan and Yu, Bowen and Yang, Jianxin and Yu, Le and Zhou, Jingren and Lin, Junyang},
  journal = {arXiv preprint arXiv:2509.17765},
  year = {2025},
  url = {https://arxiv.org/abs/2509.17765},
  eprint = {2509.17765},
  archivePrefix = {arXiv}
}

@article{lu2025desta25audio,
  title = {{DeSTA2.5-Audio}: Toward General-Purpose Large Audio Language Model with Self-Generated Cross-Modal Alignment},
  author = {Lu, Ke-Han and Chen, Zhehuai and Fu, Szu-Wei and Yang, Chao-Han Huck and Huang, Sung-Feng and Yang, Chih-Kai and Yu, Chee-En and Chen, Chun-Wei and Chen, Wei-Chih and Huang, Chien-yu and Lin, Yi-Cheng and Lin, Yu-Xiang and Fu, Chi-An and Kuan, Chun-Yi and Ren, Wenze and Chen, Xuanjun and Huang, Wei-Ping and Hu, En-Pei and Lin, Tzu-Quan and Wu, Yuan-Kuei and Huang, Kuan-Po and Huang, Hsiao-Ying and Chou, Huang-Cheng and Chang, Kai-Wei and Chiang, Cheng-Han and Ginsburg, Boris and Wang, Yu-Chiang Frank and Lee, Hung-yi},
  journal = {arXiv preprint arXiv:2507.02768},
  year = {2025},
  url = {https://arxiv.org/abs/2507.02768},
  eprint = {2507.02768},
  archivePrefix = {arXiv}
}

@inproceedings{hu2022lora,
  title = {{LoRA}: Low-Rank Adaptation of Large Language Models},
  author = {Hu, Edward J. and Shen, Yelong and Wallis, Phillip and Allen-Zhu, Zeyuan and Li, Yuanzhi and Wang, Shean and Wang, Lu and Chen, Weizhu},
  booktitle = {International Conference on Learning Representations},
  year = {2022},
  url = {https://openreview.net/forum?id=nZeVKeeFYf9},
  eprint = {2106.09685},
  archivePrefix = {arXiv}
}

@inproceedings{delcroix2018speakerbeam,
  title = {Single Channel Target Speaker Extraction and Recognition with Speaker Beam},
  author = {Delcroix, Marc and Zmolikova, Katerina and Kinoshita, Keisuke and Ogawa, Atsunori and Nakatani, Tomohiro},
  booktitle = {2018 IEEE International Conference on Acoustics, Speech and Signal Processing},
  pages = {5554--5558},
  year = {2018},
  organization = {IEEE},
  doi = {10.1109/ICASSP.2018.8462661},
  url = {https://doi.org/10.1109/ICASSP.2018.8462661}
}

@misc{delcroix2020speakerbeam,
  title = {Improving Speaker Discrimination of Target Speech Extraction with Time-Domain SpeakerBeam},
  author = {Delcroix, Marc and Ochiai, Tsubasa and Zmolikova, Katerina and Kinoshita, Keisuke and Tawara, Naohiro and Nakatani, Tomohiro and Araki, Shoko},
  year = {2020},
  eprint = {2001.08378},
  archivePrefix = {arXiv},
  primaryClass = {eess.AS},
  url = {https://arxiv.org/abs/2001.08378}
}

@inproceedings{ge2020spexplus,
  title = {{SpEx+}: A Complete Time Domain Speaker Extraction Network},
  author = {Ge, Meng and Xu, Chenglin and Wang, Longbiao and Chng, Eng Siong and Dang, Jianwu and Li, Haizhou},
  booktitle = {Proceedings of Interspeech 2020},
  pages = {1406--1410},
  year = {2020},
  doi = {10.21437/Interspeech.2020-1397},
  url = {https://doi.org/10.21437/Interspeech.2020-1397},
  eprint = {2005.04686},
  archivePrefix = {arXiv}
}

@article{wang2023tfgridnet,
  title = {{TF-GridNet}: Integrating Full- and Sub-Band Modeling for Speech Separation},
  author = {Wang, Zhong-Qiu and Cornell, Samuele and Choi, Shukjae and Lee, Younglo and Kim, Byeong-Yeol and Watanabe, Shinji},
  journal = {IEEE/ACM Transactions on Audio, Speech, and Language Processing},
  volume = {31},
  pages = {3221--3236},
  year = {2023},
  doi = {10.1109/TASLP.2023.3304482},
  url = {https://doi.org/10.1109/TASLP.2023.3304482},
  eprint = {2211.12433},
  archivePrefix = {arXiv}
}

@inproceedings{jang2017categorical,
  title = {Categorical Reparameterization with {Gumbel-Softmax}},
  author = {Jang, Eric and Gu, Shixiang and Poole, Ben},
  booktitle = {International Conference on Learning Representations},
  year = {2017},
  url = {https://openreview.net/forum?id=rkE3y85ee}
}

@inproceedings{perez2018film,
  title = {{FiLM}: Visual Reasoning with a General Conditioning Layer},
  author = {Perez, Ethan and Strub, Florian and de Vries, Harm and Dumoulin, Vincent and Courville, Aaron},
  booktitle = {Proceedings of the AAAI Conference on Artificial Intelligence},
  year = {2018},
  url = {https://ojs.aaai.org/index.php/AAAI/article/view/11671}
}

@book{moran2014phoible,
  title = {{PHOIBLE} Online},
  editor = {Moran, Steven and McCloy, Daniel and Wright, Richard},
  publisher = {Max Planck Institute for Evolutionary Anthropology},
  address = {Leipzig},
  year = {2014},
  url = {https://phoible.org/}
}

@inproceedings{khan2025uriel,
  title = {{URIEL}+: Enhancing Linguistic Inclusion and Usability in a Typological and Multilingual Knowledge Base},
  author = {Khan, Aditya and Shipton, Mason and Anugraha, David and Duan, Kaiyao and Hoang, Phuong H. and Khiu, Eric and Do{\u{g}}ru{\"o}z, A. Seza and Lee, En-Shiun Annie},
  booktitle = {Proceedings of the 31st International Conference on Computational Linguistics},
  pages = {6937--6952},
  year = {2025},
  publisher = {Association for Computational Linguistics},
  url = {https://aclanthology.org/2025.coling-main.463/}
}

@article{gordon2001phonation,
  title = {Phonation Types: A Cross-Linguistic Overview},
  author = {Gordon, Matthew and Ladefoged, Peter},
  journal = {Journal of Phonetics},
  volume = {29},
  number = {4},
  pages = {383--406},
  year = {2001},
  doi = {10.1006/jpho.2001.0147}
}

@article{ramus1999correlates,
  title = {Correlates of Linguistic Rhythm in the Speech Signal},
  author = {Ramus, Franck and Nespor, Marina and Mehler, Jacques},
  journal = {Cognition},
  volume = {73},
  number = {3},
  pages = {265--292},
  year = {1999},
  doi = {10.1016/S0010-0277(99)00058-X}
}

@inproceedings{panayotov2015librispeech,
  title = {{LibriSpeech}: An {ASR} Corpus Based on Public Domain Audio Books},
  author = {Panayotov, Vassil and Chen, Guoguo and Povey, Daniel and Khudanpur, Sanjeev},
  booktitle = {2015 IEEE International Conference on Acoustics, Speech and Signal Processing},
  pages = {5206--5210},
  year = {2015},
  organization = {IEEE},
  doi = {10.1109/ICASSP.2015.7178964},
  url = {https://doi.org/10.1109/ICASSP.2015.7178964}
}

@article{bu2017aishell,
  title = {{AISHELL-1}: An Open-Source Mandarin Speech Corpus and a Speech Recognition Baseline},
  author = {Bu, Hui and Du, Jiayu and Na, Xingyu and Wu, Bengu and Zheng, Hao},
  journal = {arXiv preprint arXiv:1709.05522},
  year = {2017},
  url = {https://arxiv.org/abs/1709.05522},
  eprint = {1709.05522},
  archivePrefix = {arXiv}
}

@inproceedings{barnard2014nchlt,
  title = {The {NCHLT} Speech Corpus of the South African Languages},
  author = {Barnard, Etienne and Davel, Marelie H. and van Heerden, Charl J. and de Wet, Febe and Badenhorst, Jaco},
  booktitle = {Proceedings of the Workshop on Spoken Language Technologies for Under-Resourced Languages},
  pages = {194--200},
  year = {2014},
  url = {https://www.isca-speech.org/archive/sltu_2014/barnard14_sltu.html}
}

@inproceedings{luong2016vivos,
  title = {{VIVOS}: A Free Vietnamese Speech Corpus for Automatic Speech Recognition},
  author = {Luong, Hieu-Thi and Vu, Hai-Quan},
  booktitle = {Proceedings of the 2016 Conference of the Oriental Chapter of the International Committee for the Co-ordination and Standardisation of Speech Databases and Assessment Techniques},
  pages = {1--4},
  year = {2016},
  url = {https://ailab.hcmus.edu.vn/vivos}
}

@inproceedings{ardila2020common,
  title = {Common Voice: A Massively-Multilingual Speech Corpus},
  author = {Ardila, Rosana and Branson, Megan and Davis, Kelly and Henretty, Michael and Kohler, Michael and Meyer, Josh and Morais, Reuben and Saunders, Lindsay and Tyers, Francis M. and Weber, Gregor},
  booktitle = {Proceedings of the Twelfth Language Resources and Evaluation Conference},
  pages = {4218--4222},
  year = {2020},
  publisher = {European Language Resources Association},
  url = {https://aclanthology.org/2020.lrec-1.520/}
}

@article{luo2019convtasnet,
  title = {{Conv-TasNet}: Surpassing Ideal Time--Frequency Magnitude Masking for Speech Separation},
  author = {Luo, Yi and Mesgarani, Nima},
  journal = {IEEE/ACM Transactions on Audio, Speech, and Language Processing},
  volume = {27},
  number = {8},
  pages = {1256--1266},
  year = {2019},
  doi = {10.1109/TASLP.2019.2915167},
  url = {https://doi.org/10.1109/TASLP.2019.2915167}
}

@inproceedings{subakan2021sepformer,
  title = {Attention Is All You Need in Speech Separation},
  author = {Subakan, Cem and Ravanelli, Mirco and Cornell, Samuele and Bronzi, Mirko and Zhong, Jianyuan},
  booktitle = {ICASSP 2021 -- 2021 IEEE International Conference on Acoustics, Speech and Signal Processing},
  pages = {21--25},
  year = {2021},
  organization = {IEEE},
  doi = {10.1109/ICASSP39728.2021.9413901},
  url = {https://doi.org/10.1109/ICASSP39728.2021.9413901}
}

@misc{openai2023whisperlargev3,
  title = {{Whisper Large-v3}},
  author = {{OpenAI}},
  year = {2023},
  howpublished = {\url{https://huggingface.co/openai/whisper-large-v3}}
}

@misc{maum2020voicefilter,
  title        = {{VoiceFilter}: Unofficial PyTorch Implementation of Google AI's VoiceFilter System},
  author       = {{MAUM AI}},
  year         = {2020},
  howpublished = {\url{https://github.com/maum-ai/voicefilter}}
}

\end{document}